\documentclass[aps,preprint]{revtex4}%
\usepackage{amsfonts}
\usepackage{amsmath}
\usepackage{amssymb}
\usepackage{graphicx}%
\begin{document}
\title{Observable effect of quantized cylindrical gravitational waves}
\author{Feifan He$^{a,b}$}
\author{Baocheng Zhang$^{b}$}
\email{zhangbaocheng@cug.edu.cn}
\affiliation{$^{a}$Institute of Geophysics and Geomatics, China University of Geosciences,
Wuhan 430074, China}
\affiliation{$^{b}$School of Mathematics and Physics, China University of Geosciences,
Wuhan 430074, China}
\begin{abstract}
We investigate the response of a model gravitational wave detector consisting
of two particles to the quantized cylindrical gravitational waves and obtain a
relation between the standard deviation of the distance between two particles
and the distance from the source to the detector. It is found that the quantum
effect carried by the cylindrical gravitational wave can be observed above
Planck scale even though the source is as far as the cosmological horizon. The
equation of motion for the change of the distance between two particles is
obtained when the cylindrical gravitational waves pass. It is surprising that
the dissipative term does not exist up to the first order approximation due to
cylindrical symmetry of the gravitational wave.

\end{abstract}
\maketitle

\setlength{\medmuskip}{0mu} \setlength{\thickmuskip}{0mu}

\section{Introduction}

Some physical effects such as black hole evaporation and early-universe
cosmology \cite{ck07,cr08} imply there should be a quantum theory about
gravity, but there is no experimental or observational evidence to support
that \cite{ck06,arp14,adm14} up to now. An important confirmation is to test
the hypothetical quanta of the gravitational field such as gravitons, but it
is nearly impossible to conclude in the foreseeable future \cite{fd13,kst21}.
Gravitational waves (GWs), however, can be used to examine the possible
implications for the quantization of gravity since the GWs have been directly
detected by LIGO in 2015 for the first time~\cite{aaa16}.

Recently, Parikh, Wilczek and Zahariade treated the GWs as quantum entity and
explored its implications for quantization of gravity from the perspective of
experimental observation \cite{pwz211,pwz212}. They calculated the effect of
quantized gravitational field on falling bodies, and found that the dynamics
of the separation of a pair of free falling particles is no longer
deterministic, but probabilistic, as acted on by a novel stochastic force. In
this paper, we will investigate this using the Einstein-Rosen wave \cite{er37}
by coupling it with a pair of free falling particles which is a simplified
model for the GW detector.

The Einstein-Rosen wave is an exact solution of general relativity with two
commuting Killing vectors and describes a cylindrical GW, so using it to study
the aspects of nonlinearity originating from Einstein gravity is convenient
and significant \cite{mt17}. Historically, it played an important role in
early attempts at defining the energy carried by gravitational waves
\cite{kst65,sc86}, since the energy of GWs is difficult to be described
locally due to the equivalence principle \cite{rai68,mab97,fjb40,llf75}. Thus,
the Einstein-Rosen waves as cylindrical GWs from some proper astrophysical
sources \cite{bsw20} could be observed really \cite{ww57} as they can carry
the energy with themselves. Moreover, the Einstein-Rosen wave has a nice
quantum description \cite{kk71} and its quantization coupled to massless
scalar field has been obtained \cite{gv05}. Although the quantization of
Einstein-Rosen cylindrical GWs has received a lot of attention
\cite{aa96,ap96,am00,gp97,dt99,mv00,cms98,rt96,ks98,bgv03,bgv04}, its
implication from the observable point of view has not been discussed. In this
paper, we study its possible quantum signatures from cylindrical GWs by
calculating the response of a model GW detector to the quantized gravitational
field. It is found that the signatures for the quantization of GWs contains
the information about the distance from the source to the detector which is
derived from the specific form of cylindrical GWs and cannot be obtained from
the general description for GWs.

The paper is organized as follows. In the second section, the theory of
Einstein-Rosen wave is reviewed. In particular, its quantized form is given
for the later discussion for the observable effect of cylindrical GWs. In the
third section, we use a simple detection model to study the observable effect
of the cylindrical GWs and some novel results are obtained. Finally, we give a
conclusion in the fourth section.

\section{Einstein-Rosen wave}

Consider a spacetime with the cylindrical symmetry and its metric can be
expressed with a conformally flat form \cite{kk71},
\begin{equation}
ds^{2}=e^{-\psi}\left[  e^{\gamma}\left(  -dT^{2}+dR^{2}\right)  +R^{2}%
d\theta^{2}\right]  +e^{\psi}dZ^{2}, \label{csm}%
\end{equation}
where the metric functions $\psi$ and $\gamma\ $depend only on the coordinates
$R$ and $T$. Using the vacuum Einstein field equation, it is obtained that
\begin{equation}
\frac{\partial^{2}\psi}{\partial T^{2}}-\frac{\partial^{2}\psi}{\partial
R^{2}}-\frac{1}{R}\frac{\partial\psi}{\partial R}=0, \label{wve}%
\end{equation}
which is the wave equation of physical degrees of freedom and has exactly the
same form as the wave equation of the cylindrically symmetric massless scalar
field $\psi$ evolving in Minkowskian spacetime background \cite{kk71}. This
means that the metric function $\psi$ represents cylindrical gravitational
waves or Einstein-Rosen waves. The metric function $\gamma$ can be obtained
as
\begin{equation}
\gamma=\frac{1}{2}\int_{0}^{R_{0}}dRR\left[  \left(  \frac{\partial\psi
}{\partial T}\right)  ^{2}+\left(  \frac{\partial\psi}{\partial R}\right)
^{2}\right]  . \label{wed}%
\end{equation}
This is the energy of cylindrical GWs in a ball of radius $R_{0}$, which
derives from the definition about C-energy introduced by Thorne \cite{kst65}
and a recent detailed discussion refers to Ref. \cite{bgp19}.

The solution of Eq.~(\ref{wve}) for a particular wave number $k$ can be
obtained as
\begin{equation}
\psi_{k}(R,T)=\frac{1}{\sqrt{2\hbar G}}J_{0}(kR)\left(  a(k)e^{-ikT}%
+a^{\dagger}(k)e^{ikT}\right)  , \label{mf}%
\end{equation}
where $J_{0}(kR)$ is the Bessel function of zeroth order. When the canonical
quantization is implemented, the parameters $a(k)$ and $a^{\dagger}(k)$ are
regarded as operators satisfying the commutation relations $\left[
a(k),a^{\dagger}(k^{\prime})\right]  =\hbar\delta(k,k^{\prime})$, and they can
be physically interpreted as annihilation and creation operators.

As discussed in the Introduction, the observable effect of cylindrical GWs is
considered at the place with a large distance from the source, so the
linearized form of this metric (\ref{csm}),
\begin{equation}
ds^{2}=(1-\psi)ds_{3}^{2}+(1+\psi)dZ^{2}, \label{met}%
\end{equation}
with $ds_{3}^{2}=-(1+\gamma)dT^{2}+(1+\gamma)dR^{2}+R^{2}d\theta^{2}$, is
adequate in the following discussion. It is noted that in the linearized
metric, the wave equation (\ref{wve}) still holds, but the energy function
$\gamma$ takes the asymptotic form. When $R\ \rightarrow\infty$, the energy of
gravitational waves is obtained as
\begin{equation}
\gamma_{\infty}=\int_{0}^{\infty}dkka^{\dagger}(k)a(k),
\end{equation}
by putting the Eq. (\ref{mf}) into Eq. (\ref{wed}) and taking the large $R$
limit \cite{ap96,bgv03}. This shows that the energy remains finite at large
$R$.

According to the analyses in Refs. \cite{kk71,rt96,bgv03}, the Hamiltonian of
this linearized gravity can be written as
\begin{equation}
H_{G0}=\int_{0}^{\infty}dR\left[  \frac{p_{\psi}^{2}}{2R}+\frac{R}{2}\left(
\frac{\partial\psi}{\partial R}\right)  ^{2}\right]  ,
\end{equation}
where the gauge fixing conditions $p_{\gamma}=0$ and $R=r$ are imposed.
$p_{\psi}$ and $p_{\gamma}$ are the canonical momenta conjugated to the metric
fields $\psi$ and $\gamma$, respectively. $R=r$ indicates that $R$ can be used
to measure the distance from the source to the detector. Noted that the metric
in Eq. (\ref{csm}) has used the gauge $R=r$ since in the initial expression
the term $R^{2}d\theta^{2}$ should be $r^{2}d\theta^{2}$. It is not hard to
confirm that $H_{G0}=\gamma_{\infty}$ when the expression of $p_{\psi}$ is
used. For the cylindrical GWs, there is another physically related Hamiltonian
$H_{G}=2\left(  1-e^{-H_{G0}/2}\right)  $ which describes the energy per unit
length along the symmetry axis in general relativity \cite{ap96}. $H_{G}$ is
related to the physical time $t$ which is gotten by the transformation
$t=e^{\gamma_{\infty}/2}T$. Furthermore, with time $t$, the annihilation and
creation operators can expressed as
\begin{align}
a_{E}(k,t)  &  =a(k)\exp\left[  -itE(k)e^{-H_{G0}/2}\right]  ,\nonumber\\
a_{E}^{\dagger}(k,t)  &  =a^{\dagger}(k)\exp\left[  itE(k)e^{-H_{G0}%
/2}\right]  ,
\end{align}
where $E(x)=2\left(  1-e^{-x/2}\right)  $. When the dimensional constants
$\hbar$ and $G$ are restored, $E(k)$ can be expressed as $\left(  1-e^{-4\hbar
G}\right)  /(4G)$, which gives $\frac{1}{\hbar}E(k)=k+O\left(  \hbar\right)
$. Thus, we take the first approximation $E(k)\sim k$ in the calculation
below. Substituting these equations into metric field in Eq. (\ref{mf}), we
have
\begin{equation}
\psi(R,t)=\frac{1}{\sqrt{2\hbar G}}\int_{0}^{\infty}dkJ_{0}(kR)\left[
a_{E}(k,t)+a_{E}^{\dagger}(k,t)\right]  .
\end{equation}
Define $q_{k}(t)=a_{E}(k,t)+a_{E}^{\dagger}(k,t)$, and decompose $\psi(R,t)$
into discrete modes. Thus, the metric field becomes
\begin{equation}
\psi(R,t)=\frac{1}{\sqrt{2\hbar G}}\sum_{k}J_{0}(kR)q_{k}(t), \label{mfr}%
\end{equation}
where the zeroth-order Bessel function $J_{0}(kR)$ satisfies the integral
relation, $\int_{0}^{\infty}dRRJ_{0}(kR)J_{0}(k^{\prime}R)=\frac{L}{2\pi
}\delta(k-k^{\prime})$, for the period boundary condition $k=2\pi R/L$.

Based on the discussion above, the Einstein-Hilbert action of linearized
cylindrically symmetric GWs can be written as
\begin{align}
S_{G}  &  =\frac{1}{64\pi G}\int_{t_{1}}^{t_{2}}\int_{0}^{\infty}dTdR\left(
p_{\psi}\frac{\partial\psi}{\partial T}-H_{0}\right) \nonumber\\
&  =\frac{1}{64\pi G}\int_{t_{1}}^{t_{2}}\int_{0}^{\infty}dTdR\frac{R}%
{2}\left[  \left(  \frac{\partial\psi}{\partial T}\right)  ^{2}-\left(
\frac{\partial\psi}{\partial R}\right)  ^{2}\right] \nonumber\\
&  =\frac{1}{2}m\int_{t_{1}}^{t_{2}}dt\sum_{k}\left(  (\dot{q}_{k}%
)^{2}-e^{-\gamma_{\infty}}k^{2}(q_{k})^{2}\right)  ,
\end{align}
where the dot denotes the derivative with respect to $t$. $m\equiv
\frac{e^{\frac{\gamma_{\infty}}{2}}L}{128\pi^{2}\hbar G^{2}}$ is similar to
the meaning of the mass, $p_{\psi}=R\frac{\partial\psi}{\partial T}$ is used
in the second line, and Eq. (\ref{mfr}) and the integral relation for the
zeroth-order Bessel function $J_{0}(kR)$ are used in the third line.

\section{Detection}

In this section, we consider the observable effect of the cylindrical GWs. In
this linearized metric (\ref{met}), the Riemann curvature tensor $R_{\text{
}0,R0}^{R}$ can be calculated, which gives the geodesic deviation equation,
$\frac{d^{2}l}{dt^{2}}=-R_{\text{ }0,R0}^{R}l$ with $l$ denoting the distance
between two free falling testing particles. With these, we can construct a
simple model to test the cylindrical GWs by the change of the distance between
a pair of particles with the action,
\begin{align}
S_{M}  &  =\int_{t_{1}}^{t_{2}}dt\left(  \frac{1}{2}m_{0}\dot{l}^{2}-\frac
{1}{2}m_{0}\left(  \dot{\gamma}-\dot{\psi}\right)  \dot{l}l\right) \nonumber\\
&  =\int_{t_{1}}^{t_{2}}dt\left(  \frac{1}{2}m_{0}\dot{l}^{2}+g\sum_{k}%
J_{0}(kR)\dot{q}_{k}\dot{l}l\right)  , \label{pmh}%
\end{align}
where $m_{0}$ is the mass of the particle and $g\equiv\frac{m_{0}}%
{2\sqrt{2\hbar G}}$ similar to the coupling parameter between the GWs and two
particles. In the second line of the calculation, Eq. (\ref{mfr}) is used, and
$\dot{\gamma}=0$ is imposed when the distance $R$ is large as required for the
discussion of the observable effect of cylindrical GWs. The independence of
$\gamma$ on the time means that the energy carried by the cylindrical GWs is
constant at large $R$ \cite{gv05,ap96,bgv03}. Thus, the effect of the
cylindrical GWs on the distance between two particles derives mainly from the
metric function $\psi$.

Together with the action of cylindrical GWs, we have the total action as
\begin{equation}
S=S_{G}+S_{M}. \label{tac}%
\end{equation}
Now we can calculate the physical effect. Similar to the consideration in
Refs. \cite{pwz211,pwz212}, the transition probability of the particles from
the state $\phi_{A}$ to state $\phi_{B}$ in time $t$, $P_{\psi_{\omega}}%
(\phi_{A}\rightarrow\phi_{B})=\sum_{\left\vert f\right\rangle }|\left\langle
f,\phi_{B}\right\vert U(t_{f},0)\left\vert \psi_{\omega},\phi_{A}\right\rangle
|^{2}$ where $\psi_{\omega}$ and $f$ are the initial and final gravitational
field states, is calculated in what follows. $U(t_{f},0)$ is the time evolving
operator which is related to the total Hamiltonian of the combined GWs with
particles. Due to the weak gravitational field, linearity allows us to write
the whole action as $S=\sum_{\omega}S_{\omega}$ with%
\begin{equation}
S_{\omega}=\int_{t_{1}}^{t_{2}}dt\left(  \frac{1}{2}m\dot{q}^{2}+\frac{1}%
{2}m_{0}\dot{l}^{2}-\frac{1}{2}m\omega^{2}q^{2}e^{-\gamma_{\infty}}%
+gJ_{0}(\omega R)\dot{q}\dot{l}l\right)  ,
\end{equation}
where the relativistic dispersion relation $\omega=k$ is used and the
subscript on $q_{k}$ is ignored for brevity. Then, the total canonical momenta
are introduced as $p=m\dot{q}+gJ_{0}(kR)\dot{l}l$ for the field and $\pi
=m_{0}\dot{l}+gJ_{0}(kR)\dot{q}l$ for the particle system, and the total
Hamiltonian is obtained as
\begin{equation}
H=\left(  \frac{p^{2}}{2m}+\frac{\pi^{2}}{2m_{0}}-\frac{gp\pi J_{0}l}{mm_{0}%
}\right)  \left(  1-\frac{g^{2}J_{0}^{2}l^{2}}{mm_{0}}\right)  ^{-1}+\frac
{1}{2}m\omega^{2}q^{2}e^{-\gamma_{\infty}}. \label{ths}%
\end{equation}
Thus, the time evolving operator can be calculated according to $U_{l}%
(t_{f},0)=\exp\left(  -\frac{i}{\hbar}\int Hdt\right)  $.

Inserting several complete bases of joint position eigenstates, $\int
dqdl\left\vert q,l\right\rangle \left\langle q,l\right\vert $, and calculating
the integral of the variables $q$ and $\pi$, the transition probability
becomes%
\begin{equation}
P_{\psi_{\omega}}[\phi_{A}\rightarrow\phi_{B}]=C\int\tilde{\mathcal{D}}%
l\tilde{\mathcal{D}}l^{\prime}e^{\frac{i}{\hbar}\int_{0}^{t_{f}}dt\frac{1}%
{2}m_{0}(\dot{l}^{2}-\dot{l}^{\prime2})}F_{\psi_{\omega}}[l,l^{\prime}],
\end{equation}
where the unrelated factor $C=\int dl_{i}dl_{i}^{\prime}dl_{f}dl_{f}^{\prime
}\phi_{A}^{\ast}(l_{i}^{\prime})\phi_{B}(l_{f}^{\prime})\phi_{B}^{\ast}%
(l_{f})\phi_{A}(l_{i})$. $F_{\psi_{\omega}}[l,l^{\prime}]=\left\langle
\psi_{\omega}\right\vert U_{l^{\prime}}^{\dagger}(t_{f},0)U_{l}(t_{f}%
,0)\left\vert \psi_{\omega}\right\rangle $ is the well-known Feynman-Vernon
influence functional \cite{fv63,ch94,blh99}. A further calculation (see the
appendix for the detail) gives
\begin{equation}
F_{\psi_{\omega}}[l,l^{\prime}]=F_{0\omega}[l,l^{\prime}]\left\langle
\psi_{\omega}\right\vert e^{W^{\ast}a^{\dagger}}e^{-Wa}\left\vert \psi
_{\omega}\right\rangle , \label{fvf}%
\end{equation}
where
\begin{equation}
W\equiv\frac{ig}{\sqrt{8m\hbar\omega}}\int_{0}^{t_{f}}dt\left(  X(t)-X^{\prime
}(t)\right)  J_{0}(\omega R)e^{-it\omega e^{-\gamma_{\infty}/2}},
\end{equation}
and
\begin{align}
F_{0\omega}[l,l^{\prime}]  &  =\exp[-\frac{g^{2}}{8m\hbar\omega}\int
_{0}^{t_{f}}\int_{0}^{t}dtdt^{\prime}J_{0}(\omega R_{1})J_{0}(\omega
R_{2})\nonumber\\
&  \times\left(  X(t)-X^{\prime}(t)\right)  \left(  X(t^{\prime}%
)e^{-i(t-t^{\prime})\omega}-X^{\prime}(t^{\prime})e^{i(t-t^{\prime})\omega
}\right)  ],
\end{align}
with $X(t)\equiv\frac{d^{2}}{dt^{2}}l^{2}(t)$ and $X^{\prime}(t)\equiv
\frac{d^{2}}{dt^{2}}l^{\prime2}(t)$. Note that we consider the source and the
two particles being in the same line. $R_{1}$ and $R_{2}$ are the distances
between the source and the two particles respectively, with $R_{2}-R_{1}=l$
for the distance between two particles.

The calculation above is made only for a single mode, and sum up all these
modes to derive the vacuum influence phase as
\begin{align}
&  i\Phi_{0}[l,l^{\prime}]\nonumber\\
&  =\sum_{k}i\Phi_{0\omega}[l,l^{\prime}]\nonumber\\
&  =-\frac{4\pi m_{0}^{2}G}{\hbar}\int_{0}^{t_{f}}\int_{0}^{t}dtdt^{\prime
}\int_{0}^{\infty}d\omega\lbrack\cos((t-t^{\prime})\omega)\nonumber\\
&  \times J_{0}(\omega R_{1})J_{0}(\omega R_{2})\left(  X(t)-X^{\prime
}(t)\right)  \left(  X(t^{\prime})-X^{\prime}(t^{\prime})\right)  ]\nonumber\\
&  +\frac{i4\pi m_{0}^{2}G}{\hbar}\int_{0}^{t_{f}}\int_{0}^{t}dtdt^{\prime
}\int_{0}^{\infty}d\omega\lbrack\sin((t-t^{\prime})\omega)\nonumber\\
&  \times J_{0}(\omega R_{1})J_{0}(\omega R_{2})\left(  X(t)-X^{\prime
}(t)\right)  \left(  X(t^{\prime})+X^{\prime}(t^{\prime})\right)  ],
\label{vip}%
\end{align}
where the relation $F_{0\omega}[l,l^{\prime}]=\exp[i\Phi_{0}[l,l^{\prime}]]$
is used. It is surprising that the second term is zero due to the relation
$\int_{0}^{\infty}J_{\nu}(ax)J_{\nu}(bx)\sin(cx)=0$ for the situation with
$Re[\nu]>-1$, $0<c<b-a$, and $0<a<b$. As discussed in Refs.
\cite{pwz211,pwz212}, the second term in Eq. (\ref{vip}) is related to the
dissipation during the interaction between the GW and the particles. So the
dissipation is zero in the situation we discuss due to the cylindrical
symmetry of GW. Actually, this term could exist when the $E(k)$ takes the
higher-order term, but these terms are suppressed by the higher-order power of
$\hbar$.

\begin{figure}[ptb]
\centering
\includegraphics[width=1\columnwidth]{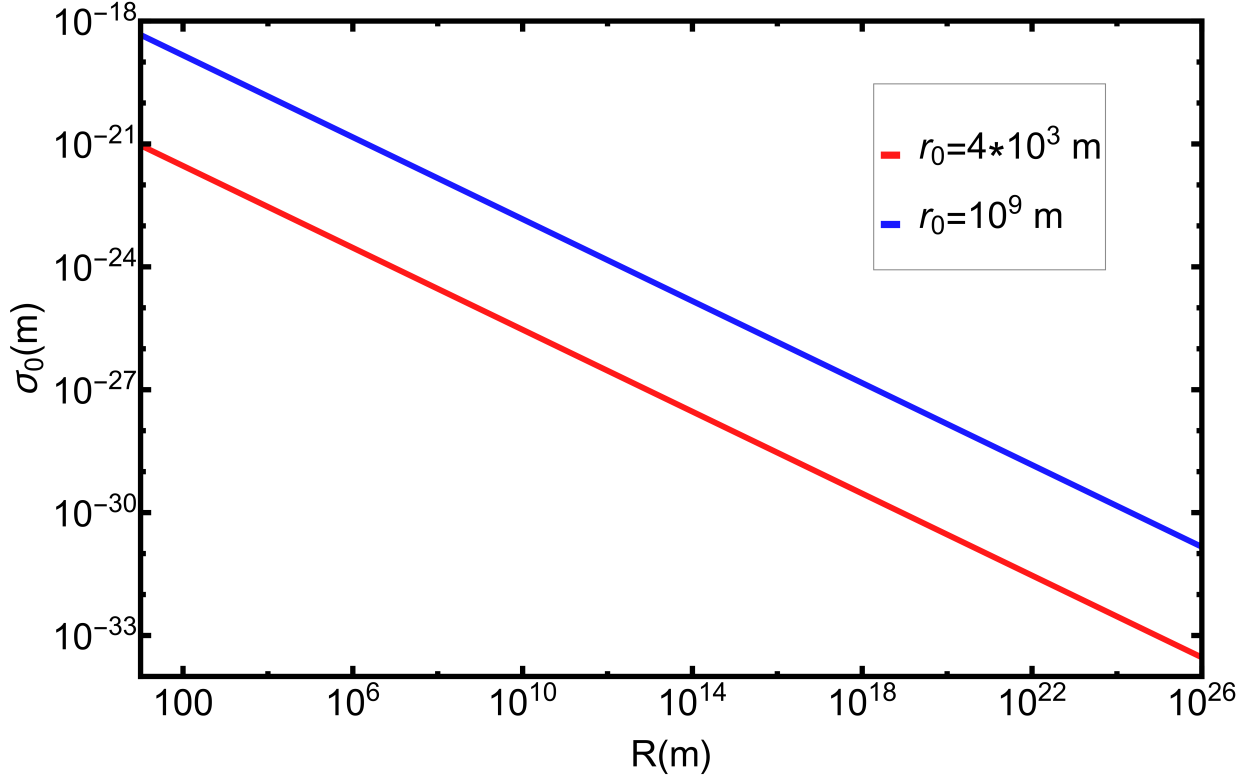} \caption{The standard
deviation $\sigma_{0}$ for the change of the length between two particles (or
test masses) influenced by cylindrical GWs as a function of the distance from
the source to the detector. The length for different lines is taken according
to the present setups or plans for GW detection, i.e. the red line represents
the arm length of ground detector like LIGO and the blue line represents the
arm length of spatial detectors like LISA, Taiji, or Tianqin. }%
\label{Fig1}%
\end{figure}

Instead of continuing to calculate the transition probability that requires an
unambiguous expression for the state $\left\vert \psi_{\omega}\right\rangle $
of the gravitational field, we use the correlation function to illustrate the
observable effect. The correlation function can be defined as in \cite{pwz212}
through the vacuum part of the influence phase by,%
\begin{equation}
A_{0}(t,t^{\prime})\equiv\frac{4\hbar G}{\pi}\int_{0}^{\infty}d\omega
\cos((t-t^{\prime})\omega)J_{0}(\omega R_{1})J_{0}(\omega R_{2}).
\end{equation}
It is noted that $A_{0}(t,t^{\prime})=$ $\langle N_{0}(t)N_{0}(t^{\prime
})\rangle=\int\mathcal{D}N_{0}\exp\left[  -\frac{1}{2}\int_{0}^{T}\int_{0}%
^{T}dtdt^{\prime}A_{0}^{-1}(t,t^{\prime})N_{0}(t)N_{0}(t^{\prime})\right]
N_{0}(t)N_{0}(t^{\prime})$ is the auto-correlation function of quantum
stochastic noise $N_{0}$ using the Feynman-Vernon trick~\cite{pwz212,fv63}. An
observable signature is obtained through the standard deviation when
$t^{\prime}=t$ given by
\begin{align}
\sigma_{0} &  =\langle\left(  l(t)-\langle l(t)\rangle\right)  ^{2}%
\rangle^{\frac{1}{2}}\nonumber\\
&  \approx\frac{l_{0}}{2}\sqrt{\langle N_{0}(t)N_{0}(t)\rangle}\nonumber\\
&  =\frac{l_{0}}{2}\sqrt{A_{0}(t,t)},\label{ssd}%
\end{align}
where $l_{0}$ is the initial distance between two particles (it can also be
considered as the arm length of the GW detector \cite{hz22}). $A_{0}%
(t,t)=\frac{4\hbar G}{\pi}\int_{0}^{\infty}d\omega J_{0}(\omega R_{1}%
)J_{0}(\omega R_{2})$ is a convergent integral, which is different from the
result in Refs. \cite{pwz211,pwz212} where the integral about $A_{0}(t,t)$ is
divergent so the frequency has to be cut off at $\omega_{max}\sim$ $2\pi
c/l_{0}$ with dipolelike approximation. A striking character of our result
(\ref{ssd}) is the dependence of the correlation function $A$ on the distance
from the source to the detector. This is demonstrated in Fig. 1. It is seen
that the samller the initial distance between two particles is, the higher
the  required sensitivity of detection would be. For the present detected
possible sources that focused on the distance from the source to the detector
about $1$ Gpc ($\sim10^{25}m$) \cite{bpa1,bpa2,bpa3}, it is found that the
present detector is unable to detect the quantum effect of GWs, since its
requirement for the ability to detect the change of $10^{-32}$m for the length
using the ground detector and $10^{-29}$m using the spatial detectors is
beyond the ability of the present technology \cite{bbf16,pas12,jl16,hw17}.
However, the quantum effect of cylindrical GW is larger than the Planck scale
even though the distance from the source to the detector reaches at the
cosmological horizon ($\sim14$Gpc), which means that the quantum property of
GW can be observed above Planck scale.

Finally, we want to give the expression for the equation of motion for the
separation of two particles. For this, such term as $\left\langle \psi
_{\omega}\right\vert e^{W^{\ast a\dagger}}e^{-Wa}\left\vert \psi_{\omega
}\right\rangle $ in Eq. (\ref{fvf}) has to be calculated. Using the same
method as in Ref. \cite{pwz212}, we have the Langevin-like equation
\begin{equation}
\ddot{l}-\frac{1}{2}\left[  \ddot{\psi}(t)+\ddot{N}_{0}(t)\right]  l(t)=0,
\label{lle}%
\end{equation}
for the state $\psi_{\omega}$ taken as the coherent state. The first term is
related to the classical effect of cylindrical GW, and the second term derives
from the quantum property of GW, which can be regarded as a stochastic noise
\cite{pwz211,pwz212} distinguished from the classically deterministic
evolution. In particular, the fifth-order derivative term that existed in the
earlier discussion \cite{pwz211,pwz212} disappears here because the second
term is zero in Eq. (\ref{vip}). Other states as the squeezed vacuum state can
be used to do the calculation, but no more new results are obtained than those
presented here or in the earlier study \cite{pwz211,pwz212}.

\section{Conclusion}

In this paper, we have investigated the quantized Einstein-Rosen wave and its
detectable effect. The Einstein-Rosen wave has the cylindrical symmetry, the
well-defined local energy, and a nice quantized form similar to that for the
quantum harmonic oscillators. Based on these, we calculated the influence of
passing cylindrical GW on a two-particle system that is a simple model for the
GW detector. Unlike the general GW detection, the parallel propagation along
the direction of two-particle connection works for our discussion. Using the
methods of the path integral and the Feynman-Vernon influence functional, we
have calculated the transition probability for the combined system of two
particles and GWs. In particular, we discuss the standard deviation for the
quantity of geodesic deviation of two-particle free motion. This can be
regarded as an observable signature. It is significant to note that the
signature carries the information about the distance from the source to the
detector. As illustrated in Fig. 1, the observable sensitivity depends not
only on the distance from the source to the detector, but also on the distance
between two particles. Interestingly, even for the sources at the cosmological
horizon, the quantum effect of cylindrical GWs could be observed above the
Planck scale. Finally, we have obtained a Langevin-like equation for the
quantity of geodesic deviation for two-particle's motion. Different from
earlier results, there is no gravitational radiation reaction term existed in
our calculation up to the first approximation due to the cylindrical symmetry
of GW. Based on these results, it is interesting to study further in what case
or how the cylindrical waves can be generated, which will be included in our
future work.

\section{Acknowledgment}

This work is supported from Grant No. 11654001 of the National Natural Science
Foundation of China (NSFC).

\section{Appendix: Derivation of equation of motion}

The purpose of the appendix is to give a detailed calculation about the
significant and different results in the third section of the main text using
the method in Ref. \cite{pwz211,pwz212}. Starting from the Hamiltonian
(\ref{ths}) of the detection model, we continue to make the calculation about
the transition probability of particle from sate $\phi_{A}$ to state $\phi
_{B}$
\begin{equation}
P_{\psi_{\omega}}(\phi_{A}\rightarrow\phi_{B})=\sum_{|f\rangle}|\langle
f,\phi_{B}|U(t_{f},0)|\psi_{\omega},\phi_{A}\rangle|^{2}, \label{FVIF}%
\end{equation}
where $|a,b\rangle\equiv|a\rangle\otimes|b\rangle$ and $U(t_{f},0)=\exp
(-\frac{i}{\hbar}\int Hdt)$ is the unitary time-evolution operator. We now
insert several complete bases of joint position eigenstates, $\int
dqdl|q,l\rangle\langle q,l|$, and have
\begin{align}
P_{\psi_{\omega}}(\phi_{A}  &  \rightarrow\phi_{B})\nonumber\\
&  =\sum_{|f\rangle}\langle\psi_{\omega},\phi_{A}|U^{\dag}(t_{f},0)|f,\phi
_{B}\rangle\langle f,\phi_{B}|U(t_{f},0)|\psi_{\omega},\phi_{A}\rangle
\nonumber\\
&  =\sum_{|f\rangle}\int dq_{i}dq_{i}^{\prime}dq_{f}dq_{f}^{\prime}%
dl_{i}dl_{i}^{\prime}dl_{f}dl_{f}^{\prime}\langle\psi_{\omega},\phi_{A}%
|q_{i}^{\prime},l_{i}^{\prime}\rangle\langle q_{i}^{\prime},l_{i}^{\prime
}|U^{\dag}(t_{f},0)|q_{f}^{\prime},l_{f}^{\prime}\rangle\nonumber\\
&  \times\langle q_{f}^{\prime},l_{f}^{\prime}|f,\phi_{B}\rangle\langle
f,\phi_{B}|q_{f},l_{f}\rangle\langle q_{f},l_{f}|U(t_{f},0)|q_{i},l_{i}%
\rangle\langle q_{i},l_{i}|\psi_{\omega},\phi_{A}\rangle\nonumber\\
&  =\int dq_{i}dq_{i}^{\prime}dq_{f}dq_{f}^{\prime}dl_{i}dl_{i}^{\prime}%
dl_{f}dl_{f}^{\prime}\psi_{\omega}^{\ast}(q_{i}^{\prime})\phi_{A}^{\ast}%
(l_{i}^{\prime})\phi_{B}(l_{f}^{\prime})\phi_{B}^{\ast}(l_{f})\psi_{\omega
}(q_{i})\phi_{A}(l_{i})\nonumber\\
&  \times\langle q_{i}^{\prime},l_{i}^{\prime}|U^{\dag}(t_{f},0)|q_{f}%
^{\prime},l_{f}^{\prime}\rangle\langle q_{f},l_{f}|U(t_{f},0)|q_{i}%
,l_{i}\rangle, \label{p1}%
\end{align}
where $\psi_{\omega}(q)$, $\phi_{A}(l)$, $\phi_{B}(l)$ are the corresponding
wave functions in position representation for the states $|\psi_{\omega
}\rangle$, $|\phi_{A}\rangle$, $|\phi_{B}\rangle$, respectively. In order to
express each of the amplitudes in canonical path-integral form, we write the
transition probability as
\begin{equation}
P_{\psi_{\omega}}[\phi_{A}\rightarrow\phi_{B}]=\int dl_{i}dl_{i}^{\prime
}dl_{f}dl_{f}^{\prime}\phi_{A}^{\ast}(l_{i}^{\prime})\phi_{B}(l_{f}^{\prime
})\phi_{B}^{\ast}(l_{f})\phi_{A}(l_{i})\int\tilde{\mathcal{D}}l\tilde
{\mathcal{D}}l^{\prime}e^{\frac{i}{\hbar}\int_{0}^{t_{f}}dt\frac{1}{2}%
m_{0}(\dot{l}^{2}-\dot{l}^{\prime2})}F_{\psi_{\omega}}[l,l^{\prime}]
\label{ta}%
\end{equation}
where the Feynman-Vernon influence functional is introduced according to the
definition as
\begin{equation}
F_{\psi_{\omega}}[l,l^{\prime}]=\langle\psi_{\omega}|U^{\dag}(t_{f}%
,0)U(t_{f},0)|\psi_{\omega}\rangle. \label{FVIF-1}%
\end{equation}
The influence functional indicates the effect of the quantized gravitational
field mode on the arm length of the detector.

In order to calculate further the Feynman-Vernon functional, we require that
we change the Hamiltonian form (\ref{ths}) in main text. Using the amplitudes
in canonical path-integral form,
\begin{equation}
\langle q_{f},l_{f}|U(t_{f},0)|q_{i},l_{i}\rangle=\int\mathcal{D}%
\pi\mathcal{D}l\mathcal{D}p\mathcal{D}q\exp{\left(  \frac{i}{\hbar}\int
_{0}^{t_{f}}dt\left(  \pi\dot{l}+p\dot{q}-H(q,p,l,\pi)\right)  \right)  },
\end{equation}
and then performing the path integral over $\pi$, we find
\begin{equation}
\langle q_{f},l_{f}|U(t_{f},0)|q_{i},l_{i}\rangle=\int\tilde{\mathcal{D}%
}le^{\frac{i}{\hbar}\int dt\frac{1}{2}m_{0}\dot{l}^{2}}\int\mathcal{D}%
p\mathcal{D}q\exp{\left(  \frac{i}{\hbar}\int_{0}^{t_{f}}dt\left(  p\dot
{q}-H_{l}(q,p)\right)  \right)  ,}%
\end{equation}
where
\begin{equation}
H_{l}(p,q)=\frac{\left(  p-gJ_{0}(\omega R)l\dot{l}\right)  ^{2}}{2m}+\frac
{1}{2}m\omega^{2}q^{2}e^{-\gamma_{\infty}}. \label{h1}%
\end{equation}
This is just the Hamiltonian required in the following calculation.
Furthermore, it can been split into a time-independent free piece and an
interaction piece, $H_{l}=H_{0}+H_{I}$ with
\begin{equation}
H_{0}=\frac{p^{2}}{2m}+\frac{1}{2}m\omega^{2}q^{2}e^{-\gamma_{\infty}},
\label{h2}%
\end{equation}%
\begin{equation}
H_{I}=-\frac{gJ_{0}(\omega R)p\dot{l}l}{m}+\frac{g^{2}J_{0}(\omega R_{1}%
)J_{0}(\omega R_{2})\dot{l}^{2}l^{2}}{2m}. \label{h3}%
\end{equation}

Notice from the form of~(\ref{h1}) that the instantaneous eigenstates are
merely those of a simple harmonic oscillator but shifted in momentum space by
$p\rightarrow p+gJ_{0}(\omega R)\dot{l}l$. Since shifts in momentum space are
generated by the position operator, we can rewrite the time-evolution operator
as
\begin{equation}
U(t_{f},0)=e^{-\frac{i}{\hbar}J_{0}(\omega R)qg\dot{l}(0)l(0)}U(t_{f}%
,0)e^{\frac{i}{\hbar}J_{0}(\omega R)qg\dot{l}(t_{f})l(t_{f})}%
\end{equation}
Using this $U(t_{f},0)$, the influence functional in Eq. (\ref{FVIF-1})
becomes
\begin{align}
&  F_{\psi_{\omega}}[l,l^{\prime}]\nonumber\\
&  =\langle\psi_{\omega}|e^{-\frac{i}{\hbar}J_{0}(\omega R)qgl^{\prime}%
(0)\dot{l}^{\prime}(0)}U^{\dag}(t_{f},0)e^{-\frac{i}{\hbar}J_{0}(\omega
R)qgl^{\prime}(t_{f})\dot{l}^{\prime}(t_{f})}e^{-\frac{i}{\hbar}J_{0}(\omega
R)qgl(t_{f})\dot{l}(t_{f})}U(t_{f},0)e^{-\frac{i}{\hbar}J_{0}(\omega
R)qgl(0)\dot{l}(0)}|\psi_{\omega}\rangle\nonumber\\
&  =\langle\psi_{\omega}|e^{-\frac{i}{\hbar}J_{0}(\omega R)q_{I}gl_{i}%
^{\prime}\dot{l}_{i}^{\prime}}U_{I}^{\dag}(t_{f})e^{-\frac{i}{\hbar}%
J_{0}(\omega R)q_{I}(t_{f})gl_{f}^{\prime}\dot{l}_{f}^{\prime}}e^{-\frac
{i}{\hbar}J_{0}(\omega R)q_{I}(t_{f})gl_{f}\dot{l}_{f}}U_{I}(t_{f}%
)e^{-\frac{i}{\hbar}J_{0}(\omega R)q_{I}(t_{f})gl_{i}\dot{l}_{i}}|\psi
_{\omega}\rangle, \label{if5}%
\end{align}
where $l_{i}=l(0)$, $l_{f}=l(t_{f})$ and quantities with a subscript $I$ are
defined in the interaction picture (e.g. $q_{I}(t)=e^{iH_{0}t/\hbar
}qe^{-iH_{0}t/\hbar}$). Since in the interaction picture, $p_{I}=m\dot{q}_{I}%
$, we can rewrite the interaction Hamiltonian as
\begin{equation}
H_{I}=J_{0}(\omega R)g\dot{q}_{I}l\dot{l}+\frac{J_{0}(\omega R_{1}%
)J_{0}(\omega R_{2})\left(  gl\dot{l}\right)  ^{2}}{2m}.
\end{equation}
Since the commutator $\left[  H_{I}(t),H_{I}(t^{\prime})\right]  =g^{2}%
J_{0}(\omega R_{1})J_{0}(\omega R_{2})l(t)\dot{l}(t)l(t^{\prime})\dot
{l}(t^{\prime})\left[  \dot{q}_{I}(t),\dot{q}_{I}(t^{\prime})\right]  $ is
easy to be confirmed to be a constant, we can eliminate the time-ordering
symbol in the interaction evolution operator $U_{I}(t_{f})=\mathcal{T}%
(e^{-\frac{i}{\hbar}\int_{0}^{t_{f}}H_{I}dldt})$ at the expense of an
additional term in the exponent which can be seen in the following form,
\begin{align}
U_{I}(t_{f})  &  =\exp\left(  -\frac{i}{\hbar}\int_{0}^{t_{f}}H_{I}dt-\frac
{1}{2\hbar^{2}}\int_{0}^{t_{f}}\int_{0}^{t}dtdt^{\prime}\left[  H_{I}%
(t),H_{I}(t^{\prime})\right]  \right) \nonumber\\
&  =\exp\left(  \frac{ig}{\hbar}\int_{0}^{t_{f}}J_{0}(\omega R)\dot{q}%
_{I}l(t)\dot{l}(t)dt+\frac{ig^{2}}{2m\hbar}\int_{0}^{t_{f}}J_{0}(\omega
R_{1})J_{0}(\omega R_{2})(l(t)\dot{l}(t))^{2}dt\right) \nonumber\\
&  \times\exp\left(  -\frac{g^{2}}{2\hbar^{2}}\int_{0}^{t_{f}}\int_{0}%
^{t}dtdt^{\prime}J_{0}(\omega R_{1})J_{0}(\omega R_{2})l(t)\dot{l}%
(t)l(t^{\prime})\dot{l}(t^{\prime})\left[  \dot{q}_{I}(t),\dot{q}%
_{I}(t^{\prime})\right]  \right)  .
\end{align}
After repeated use of integration by parts to remove the time derivatives from
the $q_{I}$ operators, this expression becomes
\begin{align}
U_{I}(t_{f})  &  =\exp\left(  \frac{ig}{\hbar}\int_{0}^{t_{f}}J_{0}(\omega
R)\dot{q}_{I}l(t)\dot{l}(t)dt+\frac{ig^{2}}{2m\hbar}\int_{0}^{t_{f}}%
J_{0}(\omega R_{1})J_{0}(\omega R_{2})(l(t)\dot{l}(t))^{2}dt\right)
\nonumber\\
&  \times\exp\left(  -\frac{g^{2}}{2\hbar^{2}}\int_{0}^{t_{f}}\int_{0}%
^{t}dtdt^{\prime}J_{0}(\omega R_{1})J_{0}(\omega R_{2})l(t)\dot{l}%
(t)l(t^{\prime})\dot{l}(t^{\prime})\left[  \dot{q}_{I}(t),\dot{q}%
_{I}(t^{\prime})\right]  \right)  .\nonumber\\
&  =\exp\left(  \frac{ig}{2\hbar}\int_{0}^{t_{f}}dtJ_{0}(\omega R)q_{I}%
(t)X(t)-\frac{ig}{\hbar}J_{0}(\omega R)q_{I}(T)l_{f}\dot{l}_{f}+\frac
{ig}{\hbar}J_{0}(\omega R)q_{I}l_{i}\dot{l}_{i}\right) \nonumber\\
&  \times\exp\left(  -\frac{g^{2}}{8\hbar^{2}}\int_{0}^{t_{f}}\int_{0}%
^{t}dtdt^{\prime}J_{0}(\omega R_{1})J_{0}(\omega R_{2})\left[  q_{I}%
(t),q_{I}(t^{\prime})\right]  X(t)X(t^{\prime})\right. \nonumber\\
&  \left.  -\frac{g^{2}}{4\hbar^{2}}\int_{0}^{t_{f}}dtJ_{0}(\omega R_{1}%
)J_{0}(\omega R_{2})[q_{I}(t),q_{I}(t^{\prime})]l_{i}\dot{l}_{i}X(t)\right.
\nonumber\\
&  \left.  +\frac{g^{2}}{4\hbar^{2}}\int_{0}^{t_{f}}dt^{\prime}J_{0}(\omega
R_{1})J_{0}(\omega R_{2})[q_{I}(t),q_{I}(t^{\prime})]l_{f}\dot{l}%
_{f}X(t^{\prime})+\frac{g^{2}}{2\hbar^{2}}J_{0}(\omega R_{1})J_{0}(\omega
R_{2})[q_{I}(t),q_{I}(t^{\prime})]l_{i}\dot{l}_{i}l_{f}\dot{l}_{f}\right)
\label{ih}%
\end{align}
where $q=q_{I}(0)$, $X(t)$ and $X^{{\prime}}(t)$ are defined after Eq.
(\ref{fvf}) in the main text.

Then, using the relation $e^{A}e^{B}=e^{A+B}e^{\frac{1}{2}[A,B]}$ where $A$
and $B$ are operators, $U_{I}(t_{f})$ can be reduced to be
\begin{align}
U_{I}(t_{f})  &  =e^{-\frac{ig}{\hbar}J_{0}(\omega R)q_{I}(t_{f})l_{f}\dot
{l}_{f}}e^{\frac{ig}{2\hbar}\int_{0}^{t_{f}}dtJ_{0}(\omega R)q_{I}%
(t)X(t)}e^{\frac{ig}{\hbar}J_{0}(\omega R)q_{I}l_{i}\dot{l}_{i}}\nonumber\\
&  \times e^{-\frac{g^{2}}{8\hbar^{2}}\int_{0}^{t_{f}}\int_{0}^{t}%
dtdt^{\prime}J_{0}(\omega R_{1})J_{0}(\omega R_{2})\left[  q_{I}%
(t),q_{I}(t^{\prime})\right]  X(t)X(t^{\prime})},
\end{align}
With this expression, we can simplify the form of the influence functional
(\ref{if5}) as
\begin{equation}
F_{\psi_{\omega}}[l,l^{\prime}]=e^{\mathcal{S}}\langle\psi_{\omega}%
|e^{-\frac{ig}{2\hbar}\int_{0}^{t_{f}}dldtJ_{0}(\omega R)q_{I}(t)X^{\prime
}(t)}e^{\frac{ig}{2\hbar}\int_{0}^{t_{f}}dldtJ_{0}(\omega R)q_{I}(t)X(t)}%
|\psi_{\omega}\rangle, \label{if3}%
\end{equation}
where
\begin{equation}
\mathcal{S}\equiv\frac{g^{2}}{8\hbar^{2}}\int_{0}^{t_{f}}\int_{0}%
^{t}dldtdt^{\prime}J_{0}(\omega R_{1})J_{0}(\omega R_{2})\left[
q_{I}(t),q_{I}(t^{\prime})\right]  \left(  X^{\prime}(t)X^{\prime}(t^{\prime
})-X(t)X(t^{\prime})\right)  ,
\end{equation}
Thus, we obtain a suitable form of the influence functional as
\begin{equation}
F_{\omega}[l,l^{\prime}]=F_{0\omega}[l,l^{\prime}]\langle\psi_{\omega
}|e^{-W^{\ast}a^{\dagger}+Wa}|\psi_{\omega}\rangle.
\end{equation}
This is the formula (\ref{fvf}) in the main text.

In order to calculate the Eq. (\ref{lle}) in the main text, the concrete
quantum state for the gravitational field has to be chosen. We choose the
coherent states, $|\psi_{\omega}\rangle=|\alpha_{\omega}\rangle$, where
$\alpha_{\omega}$ is the eigenvalue of the annihilation operator $a$,
$a|\alpha_{\omega}\rangle=\alpha_{\omega}|\alpha_{\omega}\rangle$. Since the
classical cylindrical gravitational wave mode $q$ is $q_{cl}(t)=Q_{\omega}%
\cos(\omega t+\varphi_{\omega})$, the classical cylindrical gravitational wave
can be written as
\begin{equation}
\psi(t)=J_{0}(\omega R)q_{cl}(t).
\end{equation}
The influence functional becomes
\begin{align}
F_{\omega}[l,l^{\prime}]  &  =F_{0\omega}[l,l^{\prime}]e^{-W^{\ast}%
\alpha_{\omega}^{\ast}+W\alpha_{\omega}}\nonumber\\
&  =F_{0\omega}[l,l^{\prime}]\exp\left[  \frac{ig}{2\hbar}J_{0}(\omega
R)Q_{\omega}\cos(\omega t+q_{\omega})\left(  X(t)-X^{\prime}(t)\right)
\right]  .
\end{align}
Putting all this together, we find that the transition probability can be
written as
\begin{align}
P(\phi_{A}  &  \rightarrow\phi_{B})=\int dl_{i}dl_{i}^{\prime}dl_{f}%
dl_{f}^{\prime}\phi_{A}^{\ast}(l_{i}^{\prime})\phi_{B}(l_{f}^{\prime})\phi
_{B}^{\ast}(l_{f})\phi_{A}(l_{i})\nonumber\\
&  \times\int\tilde{\mathcal{D}}l\tilde{\mathcal{D}}l^{\prime}\exp\left[
-\frac{1}{2}\int_{0}^{t_{f}}\int_{0}^{t_{f}}dtdt^{\prime}A_{0}^{-1}%
N_{0}(t)N_{0}(t^{\prime})\right] \nonumber\\
&  \times\exp\left[  \frac{i}{\hbar}\int_{0}^{t_{f}}dt\{\frac{1}{2}%
m_{0}\left(  \dot{l}^{2}-\dot{l}^{\prime2}\right)  +\frac{1}{4}m_{0}\left(
\psi(R,t)+N_{0}(t)\right)  \left(  X(t)-X^{\prime}(t)\right)  \}\right]  .
\end{align}
Using the saddle point approximation, we get the equation of motion for the
separation distance $l$ as
\begin{equation}
\frac{\partial L}{\partial l}-\frac{d}{dt}\frac{\partial L}{\partial\dot{l}%
}+\frac{d^{2}}{dt^{2}}\frac{\partial L}{\partial\ddot{l}}=0,
\end{equation}
with the Lagrangian as
\begin{equation}
L=\frac{1}{2}m_{0}\left(  \dot{l}^{2}-\dot{l}^{\prime2}\right)  +\frac{1}%
{4}m_{0}\left(  \psi(R,t)+N_{0}(t)\right)  \left(  X(t)-X^{\prime}(t)\right)
.
\end{equation}
Thus, we have the Langevin-like equation
\begin{equation}
\ddot{l}-\frac{1}{2}\left[  \ddot{N}_{0}(t)+\ddot{\psi}(R,t)\right]  l(t)=0.
\end{equation}
This is the formula (\ref{lle}) in the main text.


\begin{thebibliography}{99}
\bibitem {hz22}For the detector with the construction as LIGO, the
interference is changed like this: The length of the arm parrallel to the
propagation of GW is changed according to the way described in the text, which
is different from the change for the length of the other arm since in the case
the arm is perpendicular to the propagation of GW and there is $R_{1}=R_{2}$
in the function $A_{0}(t,t)$.
\end{thebibliography}
\end{document}